\documentclass[12pt, draftclsnofoot, onecolumn]{IEEEtran} 

\usepackage{amsmath}
\usepackage{amssymb}
\usepackage{amsfonts}
\usepackage{dsfont}
\usepackage{graphicx}
\usepackage{subcaption}
\usepackage{psfrag}
\usepackage{paralist}
\usepackage[nospace,noadjust]{cite}
\usepackage{multirow}
\usepackage{algpseudocode}
\usepackage{algorithm}
\usepackage{color}
\usepackage{amsthm}
 \usepackage[draft]{changes}
\theoremstyle{remark}
\newtheorem{remark}{Proposition}
\usepackage{ulem}
\usepackage{censor}

\colorlet{Changes@Color}{red}

\begin{document}

\title{Multi-Agent Reinforcement Learning for Energy Harvesting Two-Hop Communications with a Partially Observable State}

\author{Andrea~Ortiz,~\IEEEmembership{Student~Member,~IEEE,}        
        Tobias~Weber,~\IEEEmembership{Senior~Member,~IEEE,}
        and~Anja~Klein,~\IEEEmembership{Member,~IEEE}
\thanks{This work was funded by the LOEWE Priority Program NICER.}
}


\maketitle

\begin{abstract}
We consider an energy harvesting (EH) transmitter communicating with a receiver through an EH relay.
The harvested energy is used for data transmission, including the circuit energy consumption.
As in practical scenarios, the system's state, comprised by the 
 harvested energy, battery levels, data buffer levels, and channel gains, is only partially observable by the EH nodes. 
Moreover, the EH nodes have only outdated knowledge regarding the channel gains for their own transmit channels.
Our goal is to find distributed transmission policies aiming at maximizing the throughput. 
A channel predictor based on a Kalman filter is implemented in each EH node to estimate the current channel gain for its own channel. Furthermore, to overcome the partial observability of the system's state, the EH nodes cooperate with each other to obtain information about their parameters during a signaling phase.
We model the problem as a Markov game and propose a multi-agent reinforcement learning algorithm to find the transmission policies. 
We show the trade-off between the achievable throughput and the signaling required, and provide convergence guarantees for the proposed algorithm.
Results show that even when the signaling overhead is taken into account, the proposed algorithm outperforms other approaches that do not consider cooperation.
\end{abstract}


\begin{IEEEkeywords}
Two-hop communications, energy harvesting, decode and forward, multi-agent reinforcement learning, linear function approximation.
\end{IEEEkeywords}


%
\IEEEpeerreviewmaketitle

\section{Introduction}
\label{sec.Intro}
\IEEEPARstart{W}ireless communication nodes play an important role in many applications of wireless sensor networks such as health monitoring, surveillance or intelligent buildings. However, depending on the specific application, charging or replacing the batteries of the wireless communication nodes can be too expensive or sometimes infeasible \cite{Dargie2010}, e.g., when the nodes 
are located inside the human body, in remote locations or even inside structures. 
In order to provide sustainable service or to reduce the operating expenses, energy harvesting (EH) has been considered as a promising energy source for such wireless communication nodes.
In EH wireless communication networks, the EH capability of the nodes increases the network lifetime and can lead to perpetual operation because the nodes can use the harvested energy to recharge their batteries \cite{Ulukus2015, Ku2016}.
However, the benefits of EH are not limited to an increased network lifetime. 
The fact that the EH nodes can collect energy from natural or man-made sources, e.g., solar, chemical or electromagnetic radiation, helps to reduce greenhouse gas emissions. 
Furthermore, since the EH nodes can work independently of the power grid, EH wireless communication networks can be deployed in areas that are usually hard to reach. 
In this paper, we address the problem of how to efficiently use the harvested energy and we tackle the problem from a communications perspective, i.e., we discuss how to efficiently transmit data using the harvested energy as the only energy source.

In an EH scenario, the communication range depends on the amount of harvested energy at the EH transmitter. 
This amount of harvested energy varies according to the energy source that is considered.
For example, for energy harvesting based on electromagnetic radiation, the power density is in the order of fractions of $\text{nW}/\text{cm}^2$, and for solar energy, it is in the order of hundreds of $\text{mW}/\text{cm}^2$.
To increase the limited communication range in an EH communication scenario, relaying techniques can be considered since they are cost effective solutions for increasing the 
coverage,
throughput and robustness of wireless networks \cite{Gunduz2013, Yilmaz2010}.
By using relaying techniques, the communication between a transmitter and a receiver which are located far apart can be achieved by introducing one or more intermediate relays for reducing the communication range of each hop.
The reduction of the communication range implies a reduction of the amount of energy required for data transmission in each hop.
We focus on the case where only a single EH relay is used to assist the communication between an EH transmitter and a receiver, i.e., EH two-hop communications.
This scenario is the essential building block of more complicated EH multi-hop communication networks and exhibits all important challenges that need to be addressed when using relaying techniques, i.e., the design of transmission policies for the EH transmitter and the EH relay considering the amount of energy that is available to each of them.
Our goal is to design transmission policies aiming at an efficient use of the harvested energy at the transmitter and at the relay in order to maximize the throughput. This problem is equivalent to the minimization of the time required to transmit a given amount of data \cite{Orhan2012}.

\subsection{Related Work}
The study of EH wireless communications has been based on three different approaches, namely, offline approaches \cite{Tutuncuoglu2012, Ozel2011, Orhan2013, Orhan2012,
Orhan2012ProcCost, Li2017, OzelBC2012, Fu2016, Yang2012, Wang2015, Gunduz2011, Orhan2015, Ortiz2015, LuoYaming2013, Varan2013, 
Tutuncuoglu2013, Varan2013b}, online approaches \cite{Lei2009, Luo2013, Zeng2015, Minasian2014, Amirnavaei2015} and learning approaches 
\cite{Blasco2013, Ortiz2016, Gregori2016, Xiao2015, Chaoming2009, Shresthamali2017, Blasco2015, Ortiz20162hop}.
The offline approaches assume complete non-causal knowledge regarding the EH, the data arrival and the channel fading processes. 
Although this assumption cannot be fulfilled in reality, the offline approaches are useful to derive upper bounds of the performance.
A more relaxed assumption is considered by the online approaches where only statistical knowledge is assumed to be available in advance. 
However, in real scenarios this statistical knowledge might not be available, especially if non-stationary EH, data arrival and channel fading processes are 
considered.
The requirement of complete non-causal knowledge (offline approaches) or statistical knowledge (online approaches) can be overcome if a learning approach is 
considered. 
This is because in learning approaches, more specifically in reinforcement learning (RL), an agent learns how to behave in an unknown 
environment by interacting with it. 
For EH communications, the agent can be the EH transmitter and the environment is the unknown EH, data arrival and channel fading processes.
In the following, we give an overview of the state of the art of these approaches, first for EH point-to-point communications and secondly for EH two-hop communications.

\subsubsection{Offline approaches for EH  point-to-point communications}
Offline EH point-to-point communications have been investigated in \cite{Tutuncuoglu2012, Ozel2011,Orhan2013, Orhan2012ProcCost, Li2017}. 
In \cite{Tutuncuoglu2012}, it is shown that the throughput maximization problem within a deadline is equivalent to the minimization of the completion time given that a fixed amount of data needs to be transmitted. 
A similar scenario is investigated in \cite{Ozel2011}, where the authors consider a fading channel between the transmitter and the receiver, 
and a modified water-filling algorithm is proposed to maximize the throughput within a deadline. 
In \cite{Orhan2013Dist}, the minimization of the distortion of the received messages considering that each message has to be reconstructed at the destination within a certain deadline is studied.
Additionally, the processing costs at the transmitter in a point-to-point scenario are analyzed in \cite{Orhan2012ProcCost}. 
In \cite{Li2017}, we consider the case where each data packet to be sent has an individual deadline. 
For this scenario, we formulate optimization problems to consider the delay-constrained throughput maximization problem as well as the delay-constrained energy 
minimization problem.
\subsubsection{Online approaches for EH  point-to-point communications}
Online approaches for point-to-point scenarios are investigated in \cite{Lei2009, Luo2013}. 
In \cite{Lei2009}, statistical information about the distribution of the importance of the messages is assumed, and an on-off mechanism at the transmitter is 
considered.
A save-then-transmit protocol for system outage minimization is considered in \cite{Luo2013} where it is assumed that a fixed amount of data needs to be 
transmitted during each time interval.
\subsubsection{Learning approaches for EH  point-to-point communications}
Learning approaches have been applied to EH point-to-point scenarios in \cite{Blasco2013, Ortiz2016, Gregori2016, Xiao2015, Chaoming2009, Shresthamali2017}. 
In \cite{Blasco2013}, the well-known RL algorithm Q-learning is applied to maximize the throughput in a fixed period of time. 
In our previous work \cite{Ortiz2016}, we combine the RL algorithm State-Action-Reward-State-Action (SARSA) with linear function 
approximation to enable the use of incoming energy and channel values which are taken from a continuous range. 
In \cite{Gregori2016}, the authors use online convex optimization to derive online algorithms to learn the transmission policy from previous observations. Bayesian RL is used in \cite{Xiao2015} and the authors of \cite{Shresthamali2017} use weather forecast data to enhance the performance of RL.
\subsubsection{Offline approaches for EH two-hop communications}
For EH two-hop communications, offline approaches have been the major direction of state of the art research \cite{Gunduz2011, Orhan2015,Ortiz2015, 
LuoYaming2013, Varan2013}. 
In \cite{Gunduz2011}, the throughput maximization problem within a deadline is studied and two cases are distinguished, namely a full-duplex and a half-duplex
relay. 
For the case of a full-duplex relay, an optimal transmission scheme is provided. 
However, in the half-duplex case, a simplified scenario is assumed where a single energy arrival is considered at the transmitter.
In \cite{Orhan2015}, the authors formulate a convex optimization problem to find offline transmission policies for multiple parallel relays in a decode-and-forward EH 
two-hop communication scenario.
Half-duplex amplify-and-forward EH two-hop communications are considered in our previous work \cite{Ortiz2015}. 
In this case, we used D.C. programming to find the optimal power allocation.
In \cite{LuoYaming2013}, the throughput maximization problem is investigated when the transmitter harvests energy multiple times and the 
amplify-and-forward relay has only one energy arrival.
In \cite{Varan2013}, the impact of a finite data buffer at the relay is investigated. Similar to the previous case, it is assumed that the 
transmitter harvests energy several times while the relay harvests energy only once. 
\subsubsection{Online approaches for EH  two-hop communications}
In \cite{Minasian2014} and \cite{Amirnavaei2015}, online approaches are considered.
In \cite{Minasian2014}, a half-duplex amplify-and-forward EH two-hop communications scenario is studied. 
The authors assume statistical knowledge about the energy harvesting process and find the transmission policy using discrete dynamic programming.
A similar scenario is considered in \cite{Amirnavaei2015}, where the power allocation policy is found using Lyapunov optimization techniques.
\subsubsection{Learning approaches for EH  two-hop communications}
In our previous work \cite{Ortiz20162hop}, the two-hop communications scenario is separated into two point-to-point scenarios and the transmitter and the relay solve independent 
RL problems to find the transmission policies that aim at maximizing the throughput.
In the present paper, we study the case when the transmitter and the relay cooperate with each other to overcome the partial observability of the system's state and to
improve the achievable throughput.

\subsection{Contributions}
We focus our work on EH two-hop communications.
In contrast to the state of the art, we consider a realistic scenario in which the state of the system is only partially observable to the EH nodes.
This means, in each time interval, each EH node only knows the values of its own current and past parameters, i.e., the amounts of incoming energy, the battery levels, the 
data buffer levels and the past channel gains for its own transmit channel. 
For the transmitter, this channel gain corresponds to the channel between the transmitter and the relay, and for the relay, this channel gain corresponds to the channel between the 
relay and the receiver.
We use a channel predictor based on a Kalman filter in each EH node in order to obtain a current estimate of the channel gain.
Furthermore, to overcome the partial observability of the system's state, we propose a signaling phase in which the EH 
nodes exchange information about their current parameters, as commonly done in wireless sensor networks \cite{Rajendran2006}.
We are interested in a distributed solution where each EH node finds its own transmission policy taking into account its observation of the system's state and the knowledge it has obtained during the signaling phase.
Considering that the problem consists of two agents, the transmitter and the relay, who should make simultaneous decisions to achieve a common goal, i.e., decide on the transmit powers in order to maximize the throughput, we model this scenario as a Markov game.
This is because Markov games provide a framework to include multiple decision making agents with interacting or competing goals \cite{Littman94}.
Additionally, to find the distributed transmission policies at the transmitter and at the relay, we propose a multi-agent RL algorithm.
The use of RL is motivated by the fact that complete non-causal knowledge is unavailable. 
As a consequence, standard optimization techniques cannot be used.
In addition, to validate our proposed multi-agent RL algorithm, we derive convergence guarantees based on RL assuming that the EH nodes are able to observe the system's state, i.e., when the channel prediction and the transmission of the signaling are successful, and a constant learning rate is used.
Moreover, by numerical results we show that the performance of the proposed algorithm has only a small degradation compared to the offline case which requires complete non-causal knowledge.
Additionally, we show that even when the overhead caused by the signaling phase is taken into account, the proposed algorithm outperforms other approaches that do not consider cooperation among the EH nodes, and therefore do not require a signaling phase.

\subsection{Organization of the paper}
The rest of the paper is organized as follows. In Section \ref{sec.SystemModel}, the system model is presented and the transmission scheme is explained.
In Section \ref{sec.globalKnowledge}, the EH two-hop communications problem is addressed. 
We model the problem as a Markov game and apply multi-agent RL to find the transmission policies at the transmitter and at the relay. 
Convergence guarantees for the proposed algorithm are presented in Section \ref{sec.Convergence}.
Numerical performance results are presented in Section \ref{sec.Simulation} and Section \ref{sec.Conclusion} concludes the paper.


\section{System Model}
\label{sec.SystemModel}

\begin{figure}
\centering
\includegraphics[width=0.5\textwidth]{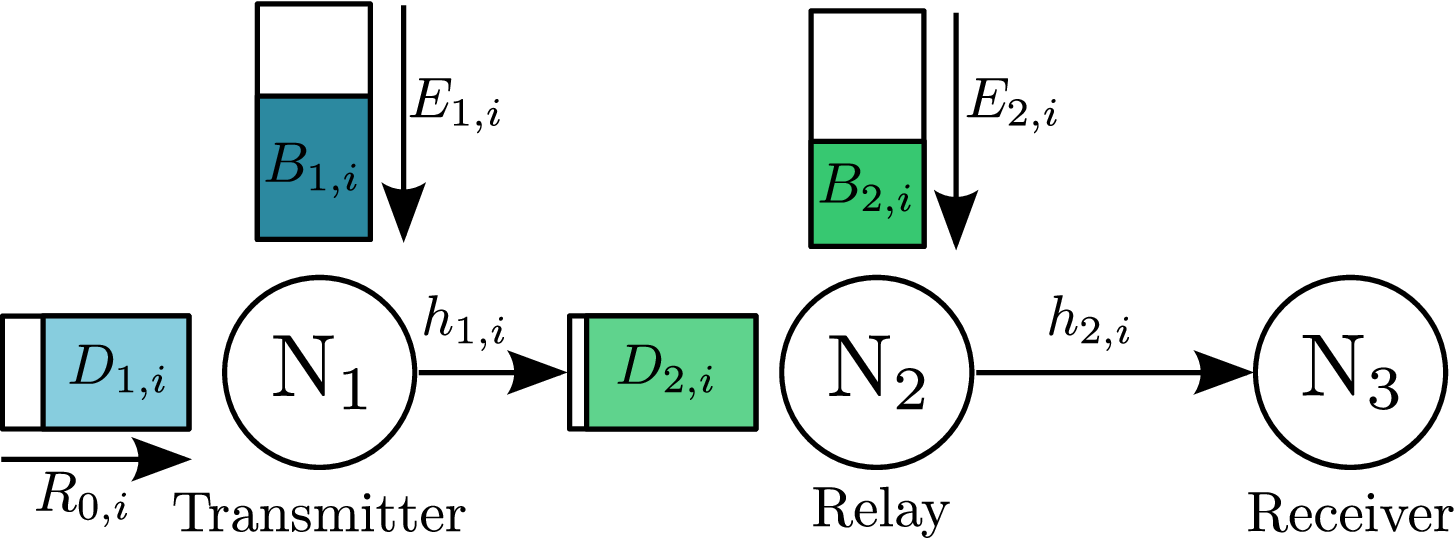}
\caption{EH two-hop communications scenario.}
\label{fig:systemModel}%
 \vspace{-2mm}
\end{figure}

An EH two-hop communication scenario consisting of three single-antenna nodes is considered. 
The symbol $\text{N}_k$, $k\in\{1,2,3\}$, is used to label the nodes. 
As depicted in Fig. \ref{fig:systemModel}, the transmitter $\text{N}_1$ wants to transmit data to the receiver $\text{N}_3$. 
It is assumed that the link between these two nodes is weak. 
Therefore, the nodes cannot communicate directly. 
To enable communications, ${\text{N}_2}$ acts as a full-duplex decode-and-forward relay. 
It is assumed that the relay $\text{N}_2$ forwards the data from ${\text{N}_1}$ to ${\text{N}_3}$ and it is able to perfectly 
cancel the self-interference caused by its 
transmission.
Furthermore, a time slotted system using $I$ time intervals is considered with a constant duration $\tau$ for each time interval $i$, $i=1,...,I$.

$\text{N}_1$ and $\text{N}_2$ harvest energy from the environment. It is assumed that this energy is used for data transmission.
Our energy consumption model includes both, the circuit energy $E_\text{Circ}$ required by all the modules that process the signal to be transmitted, e.g., base-band signal processing  unit,  digital-to-analog  converter, etc., and the transmit energy $E_\text{Tx}$ required to amplify the signal to be transmitted.
For simplicity, we assume that the energy consumed when the nodes are in sleep mode is much smaller than the energy consumed while transmitting and can be neglected \cite{Xu2014}.

We consider a discrete time model in which at the beginning of each time interval $i$, an amount of energy $E_{l,i} \in \mathbb{R}^+,\; l\in\{1,2\}$ is 
received 
by $\text{N}_l$.
The amount of energy $E_{l,i}$ may also take the value $E_{l,i}=0$ to include the case when $\text{N}_l$ does not harvest energy in time interval 
$i$.
The maximum amount of energy that can be harvested at $\text{N}_l$, termed $E_{\text{max},l}$, depends on the energy source that is used.
The harvested energy $E_{l,i}$ is stored in a rechargeable battery with maximum capacity $B_{\text{max},l}$.
It is assumed that no energy is lost in the process of storing or retrieving energy from the batteries. 
The battery levels $B_{l,i}$ are always measured at the beginning of each time interval $i$.
Furthermore, it is assumed that at the beginning of time interval $i=1$, the nodes have not yet harvested any energy and their batteries are empty, i.e., 
$B_{l,1}=0$.

The data available for transmission at $\text{N}_1$ results from its own data arrival process. 
It is assumed that at the beginning of time interval $i$, a data packet of $R_{0,i}$ bits is received by $\text{N}_1$ and the incoming data is 
stored in a finite data buffer with size $D_{\text{max},1}$, measured in 
bits.
Moreover, it is assumed that ${\text{N}_2}$  does not have any own data to transmit to the other nodes. 
Consequently, $\text{N}_2$ can only retransmit what it has received from ${\text{N}_1}$.
Similar to $\text{N}_1$, $\text{N}_2$ receives $R_{1,i}$ bits in time interval $i$ and stores them in its data buffer.
The maximum amount of data which $\text{N}_2$ can store is limited by the size of its data buffer which is given by $D_{\text{max},2}$.
The data buffer level of $\text{N}_l$ is measured at the beginning of time interval $i$ and is denoted by $D_{l,i}$.
It is assumed that at at the beginning of time interval $i=1$, both data buffers are empty, i.e., $D_{l,1}=0$.

The fading channel from $\text{N}_1$ to $\text{N}_2$ is described by the channel gain
$g_{1,i} \in \mathbb{R}$ while the 
fading channel between $\text{N}_2$ and $\text{N}_3$ is described by the channel gain
$g_{2,i} \in \mathbb{R}$.
It is assumed that the channels stay constant for the duration of one time interval.
Only outdated knowledge regarding the transmitter side channel gains is assumed. This means, that at the beginning of time interval $i$, only the channel 
gains up to time interval $i-1$ are known at $\text{N}_l$.
To obtain an estimate $\hat{g}_{l,i}$ of the channel gain in time interval $i$, $\text{N}_l$ uses a channel predictor based on a Kalman filter. 
This channel predictor is explained in Section \ref{subsec.kalman}.
The noise at $\text{N}_2$ and $\text{N}_3$ is assumed to be independent and identically distributed (i.i.d.) zero mean additive white Gaussian 
noise (AWGN) with variance $\sigma^2_2=\sigma^2_3=\sigma^2$.
Moreover, in our model the interference is treated as noise.
Additionally, the same bandwidth $W$ is assumed to be available at the EH nodes for the  transmission from $\text{N}_1$ to $\text{N}_2$ 
and from $\text{N}_2$ to $\text{N}_3$.

In this paper, we consider a signaling phase of duration $\tau_\text{sig}$ in which the EH nodes exchange information about their current parameters\footnote{As mentioned in Section \ref{sec.Intro}, a node's current parameters are the amount of incoming energy, the battery level, the data buffer level and the past observed channel gain for its own transmit channel.}.
After the signaling phase, the nodes transmit data during the rest of the time interval.
The time available for the data transmission is $\tau_\text{data}=\tau-\tau_\text{sig}$.
\begin{figure}
    \centering    
      \includegraphics[width=0.35\textwidth]{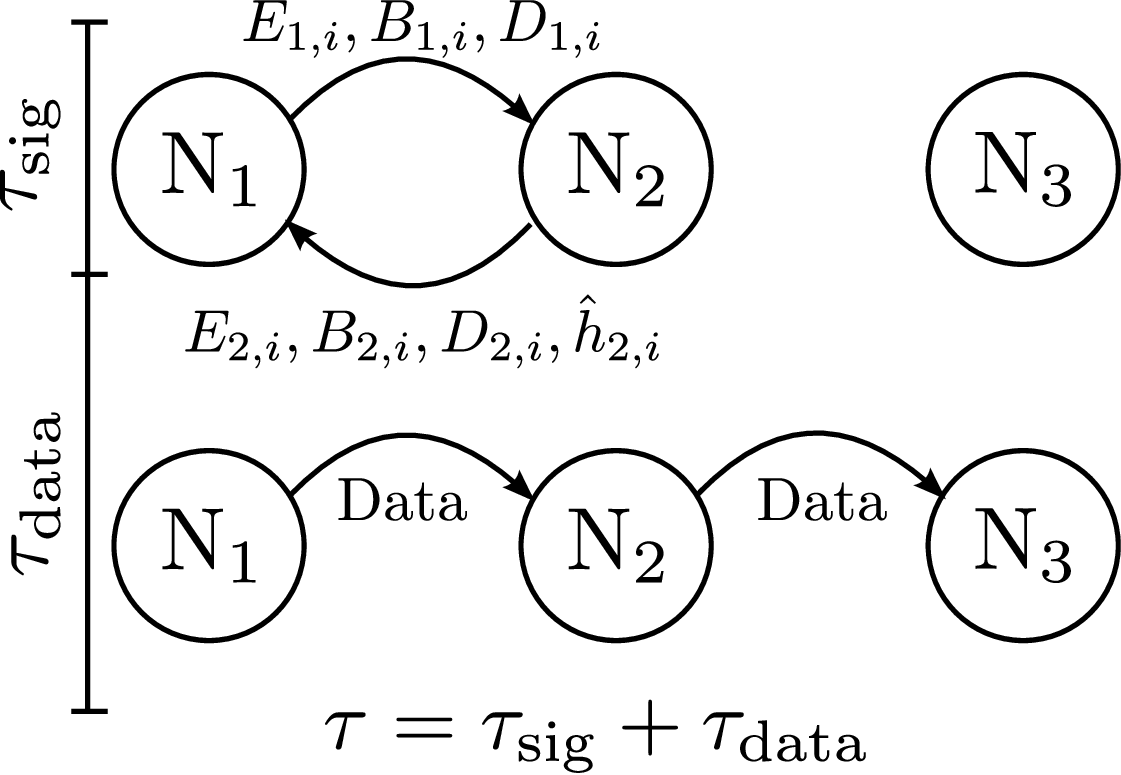}
    \caption{Communication schemes.}
    \label{fig:commSchemes}%
\end{figure}
The signaling phase provides each EH node with information about the other node's parameters as shown in Fig. \ref{fig:commSchemes}.
It should be noted that the estimated channel gain $\hat{g}_{1,i}$ does not need to be sent by $\text{N}_1$ to $\text{N}_2$. 
This is because the channel gain $g_{1,i}$ is assumed to be known at $\text{N}_2$ in time interval $i$.
To exchange their parameters, the EH nodes use a transmit power $p_{\text{sig},l,i}$ which is kept constant during $\tau_\text{sig}$. 
The duration $\tau_\text{sig}$ depends on the tolerable quantization error, the available bandwidth and the channel gains, and 
it is explained in detail in Section \ref{subsec.signaling}.
When $\text{N}_l$ does not have enough energy to transmit during the signaling phase, i.e., $B_{l,i}<\tau_\text{sig}p_{\text{sig},l,i}$, 
$p_{\text{sig},l,i}$ is set to zero and the node does not transmit anything during $\tau_\text{sig}$.
In every time interval, after $\tau_\text{sig}$, $\text{N}_1$ and $\text{N}_2$ decide independently on the transmit power $p_{l,i}$ to be used for the transmission of data which is kept constant during $\tau_\text{data}$ \cite{Tutuncuoglu2012}.
The throughput achieved by $\text{N}_l$ {in time interval $i$} is the amount of data received by $\text{N}_{l+1}$ which is 
measured in bits and is given by
\begin{equation}
 R_{l,i}=\tau_\text{data}W\log_2\left(1+\frac{|g_{l,i}|^2p_{l,i}}{\sigma^2}\right).
 \label{eq:throughput}
\end{equation}

Only the energy already stored in the battery can be used for the transmission of either the signaling or the data. 
As a result, the energy causality constraint $E_\text{Circ}+\tau_\text{sig}p_{\text{sig},l,i}+\tau_\text{data}p_{l,i}\leq B_{l,i}$ has to be fulfilled.
Moreover, in the selection of $p_{l,i}$ the finite capacities of the batteries have to be considered and battery overflow situations, in which part of the 
harvested energy 
is lost because the batteries are full, should be avoided. The battery overflow constraint is given by
\begin{equation}
  B_{l,i} + E_{l,i} - E_\text{Circ} - \tau_\text{data}p_{l,i} -\tau_\text{sig}p_{\text{sig},l,i} \leq B_{\text{max},l}.
  \label{eq:energyOverflow}
\end{equation}

Additionally, the data arrival process at $\text{N}_l$ should be considered.
Only data already stored in the data buffer can be transmitted.
Therefore, the data causality constraint $R_{l,i}\leq D_{l,i}$
has to be fulfilled in every time interval.
Moreover, in order not to lose data, data buffer overflows should be avoided. 
However, it should be noted that this cannot always be avoided because the transmission of data depends on the available energy.
As we aim at reducing the number of data buffer overflows in order to maximize the throughput, we define the data buffer overflow condition in an 
analogous way to the battery overflow constraint in (\ref{eq:energyOverflow}) as
\begin{equation}
  D_{l,i} + R_{l-1,i} - R_{l,i} \leq D_{\text{max},l}.
  \label{eq:dataOverflow}
\end{equation}


\section{Proposed Multi-Agent Reinforcement Learning (MARL) for EH Two-Hop Communications with Partially Observable State}
\label{sec.globalKnowledge}

In this section, we model the EH two-hop communication problem as a Markov game and introduce the proposed multi-agent RL (MARL) algorithm.
The proposed algorithm is used to find transmission policies at the transmitter and at the relay aiming at maximizing the throughput when the system's state is only partially observable by the EH nodes. 
The partial observability of the system's state is due to the fact that in a given time interval, the EH nodes only know their own current and past parameters.
To overcome the partial observability of the system's state, we propose to include a Kalman filter based channel predictor in each EH node to predict the channel gain in each time interval. Additionally, we consider a signaling phase in which the nodes exchange their current parameters.

\subsection{Markov game}
\label{subsec.markovGame}
In our scenario, $\text{N}_1$ and $\text{N}_2$ independently decide on the transmit power to use for data transmission.
When only one node is considered, e.g., in a point-to-point scenario, these decision-making situations can be modeled as Markov decision processes. 
However, in our case, the achieved throughput depends on the transmission policies of both $\text{N}_1$ and $\text{N}_2$. 
Consequently, Markov decision processes are no longer suitable because more than one node has to be considered. 
Markov games are a generalization of Markov decision processes and are used to model decision-making situations in which more than one agent is involved 
\cite{Littman94}.

A Markov game composed of $n$ players is defined by the set $\mathcal{S}$ of states in which the system can be, the sets 
$\mathcal{A}_1, ..., \mathcal{A}_n$ of actions of each player, the transition function $\mathcal{T}$ and the reward functions $\mathcal{R}_1, ..., 
\mathcal{R}_n$ for each player \cite{Littman2001}.
In our case, the players correspond to $\text{N}_1$ and $\text{N}_2$. Consequently, we consider $n=2$ players. 
Each state contains the parameters of both nodes, i.e., the amounts of incoming energy $E_{l,i}$, the battery levels $B_{l,i}$, the channel gains $g_{l,i}$, and the data buffer levels $D_{l,i}$.
In other words, in time interval $i$, the corresponding state $S_i \in \mathcal{S}$ is the tuple [$E_{1,i}$, $E_{2,i}$, $B_{1,i}$, 
$B_{2,i}$, $g_{1,i}$, $g_{2,i}$, $D_{1,i}$, $D_{2,i}$]. 
The set $\mathcal{S}$ comprises an infinite number of states $S_i$ because the parameters can take any value in a continuous range.
Moreover, as $\text{N}_1$ and $\text{N}_2$ only know their own parameters and the current channel gains $g_{1,i}$ and $g_{2,i}$ are not known in time interval $i$, $S_i$ is only partially observable by each of the EH nodes.
We define the sets of actions $\mathcal{A}_1$ and $\mathcal{A}_2$ for $\text{N}_1$ and $\text{N}_2$, respectively, as finite sets in order to simplify the selection of the transmit powers values. 
The sets of actions are defined as $p_{l,i}\in \mathcal{A}_l=\{0,\delta, 2\delta, ...,B_{\text{max},l}\}$, where $\delta $ is a step size.
The transition function $\mathcal{T}$ is defined as $\mathcal{T}: \mathcal{S}\times\mathcal{A}_1\times\mathcal{A}_2\rightarrow \mathcal{S}$ and it specifies that, given state $S_i$ and selecting $p_{1,i} \in \mathcal{A}_1$ and $p_{2,i}\in \mathcal{A}_2$, the 
nodes reach state $S_{i+1}$, i.e., 
$S_{i+1}=\mathcal{T}(S_i, p_{1,i}, p_{2,i})$.
The reward function $\mathcal{R}_l$ gives the immediate reward obtained by $\text{N}_l$ when $p_{l,i}$ is selected 
while being in state $S_i$.
In our case, the nodes aim at maximizing the throughput, i.e., the amount of data received by $\text{N}_3$. 
Consequently, $\text{N}_1$ and  $\text{N}_2$ share the same objective $\mathcal{R}_1=\mathcal{R}_2=\mathcal{R}$.
Moreover,  the reward $R_{i} \in \mathcal{R}$ corresponds to the throughput achieved in time interval $i$ for $l=2$ and it is calculated using 
(\ref{eq:throughput}).
As $\text{N}_l$ has only partial causal knowledge about its state, it does not know how much energy will be harvested, how much data will arrive or how the channel will be  in future time intervals. 
We consider this uncertainty by defining the discount factor of future rewards $\gamma$, 
$0\leq\gamma\leq1$, which quantifies the preference of achieving a larger throughput in the current time interval over future ones.
Our goal is to select $p_{l,i},\, \forall l,i$, in order to maximize the expected throughput
\begin{equation}
R = \underset{{I\rightarrow\infty}}{\mathrm{lim}} \mathbb{E}\left[\sum_{i=1}^{I}{\gamma^iR_{2,i}}\right].
\label{eq:objective} 
\end{equation}
To maximize the expected throughput, we need to find the transmission policies for $\text{N}_1$ and $\text{N}_2$ which correspond to the 
transmit powers to be used for data transmission in each time interval. 
Each transmission policy $\pi_l,\, l\in\{1,2\}$ is a mapping from a given state $S_{i}$ to $p_{l,i}$ that should be 
selected by the node, i.e. $p_{l,i}=\pi_l(S_{l,i})$.

\subsection{Single-agent reinforcement learning}
\label{subsec.SARL}
To facilitate the description of the proposed MARL algorithm, let us first consider the single-agent case.
For this purpose, let us assume that an ideal central entity has, in each time interval, perfect knowledge about $S_i$ and uses RL to find the combined policy $\Pi$, with $\Pi=(\pi_1,\pi_2)$. 
$\Pi$ can be evaluated using the so-called action-value function $\text{Q}^\Pi(S_{i},{P}_{i})$, with ${P}_i=(p_{1,i},p_{2,i})$, which is defined as the 
expected 
reward starting from state $S_i$, selecting $P_i$ and following $\Pi$ thereafter \cite{Sutton1998}.
The optimal policy $\Pi^*$ is the policy whose action-value function's value is greater than or equal to the value obtained by any other policy for every 
state $S_i$ and $P_i$. 
The corresponding action-value function for the optimal policy $\Pi^*$ is denoted by $\text{Q}^*$.
Furthermore, determining $\Pi^*$ is straightforward when $\text{Q}^*$ is known because for each state $S_{i}$, any $P_{i}$ 
that maximizes $\text{Q}^*(S_{i}, P_{i})$ is an optimal action. 
The action-value function cannot be calculated in advance because only causal information is available at the nodes and the statistics of the EH, 
data arrival and channel fading processes are unknown. 
To overcome this, RL builds an estimate of the action-value function $\text{Q}^{\Pi}$. 
Specifically, we consider the temporal-difference RL algorithm State-Action-Reward-State-Action (SARSA) which builds the estimate 
based on the visited states and the obtained rewards. 
The SARSA update rule for the estimate of the action value function $\text{Q}^\Pi(S_i,P_i)$ is given by
\begin{align}
\text{Q}_{i+1}^\Pi(S_i,P_i)= \, & \text{Q}_i^\Pi(S_i,P_i)(1-\alpha_i) + \alpha_i \left[R_{i}+\gamma 
\text{Q}_i(S_{i+1},P_{i+1})\right] \label{eq:centralizedSARSA}
\end{align}
\cite{Sutton1998}, where $\alpha_i$ is a small positive fraction which influences the learning rate.
Additionally, we include the sub-index $i$ in the definition of the update to emphasize the fact that the update changes the estimated value of the 
action-value function for the pair $(S_i, P_i)$ in time interval $i$.

\subsection{Multi-agent reinforcement learning}
\label{subsec.MARL}
We propose a MARL algorithm based on the SARSA update in (\ref{eq:centralizedSARSA}). 
In our scenario, the nodes have a common objective, which is to maximize the expected throughput, 
and in every time interval they make independent decisions that aim at achieving this objective. 
As the nodes do not know in advance the transmit power which will be selected by the other node, they cannot build an estimate of the 
centralized action-value function $\text{Q}^\Pi(S_i, P_i)$.
Instead of the action-value function $\text{Q}^\Pi(S_i, P_i)$, in the proposed MARL algorithm, each node builds an estimate of its local action-value 
function $\text{q}_l^{\pi_l}(S_i, p_{l,i})$.
To select $p_{l,i}$, each node follows the $\epsilon$-greedy policy \cite{Sutton1998}, i.e., $\text{N}_l$ acts greedily with respect to its action-value function with a probability of $1-\epsilon$, 
this means
\begin{equation}
\text{Pr}\left[p_{l,i}=\underset{{p_{l,k}\in\mathcal{A}_l}}{\mathrm{max}}\:\hat{\text{q}}_l^{\pi_l}(S_{i},p_{l,k}) \right]=1-\epsilon,\quad
0<\epsilon<1.
\label{eq:egreedy}
\end{equation}
However, with a probability of $\epsilon$, $\text{N}_l$ will randomly select a transmit power value from the set $\mathcal{A}_l$.
This method provides a trade-off between the exploration of new transmit power values and the exploitation of the known ones \cite{Sutton1998, Russell2010}.

$\text{q}_l^{\pi_l}(S_i, p_{l,i})$ is a projection of the centralized $\text{Q}^\Pi(S_i,P_i)$ in which the  nodes will only 
update their current estimate of $\text{q}_{l}^{\pi_l}(S_i,p_{l,i})$ if the value of the update is larger than the current one. 
The relation between $\text{Q}^\Pi(S_i,P_i)$ and $\text{q}_{l}^{\pi_l}(S_i,p_{l,i})$ is presented in detail in Section \ref{sec.Convergence} and the proposed 
updating rule for $\text{q}_l^{\pi_l}(S_i, p_{l,i})$ is given by
\begin{align}{
\text{q}^{\pi_l}_{l,i+1}(S_i, p_{l,i}) = \text{max}}&{\left\{\text{q}^{\pi_l}_{l,i}(S_i,p_{l,i}),\, (1-\alpha_i)\text{q}^{\pi_l}_{l,i}(S_i,p_{l,i}) +\alpha_i\left[R_i + \gamma \text{q}^{\pi_l}_{l,i}(S_{i+1},p_{l,i+1})\right]\right\}.}
\label{eq:distSARSA}
\end{align}
The action-value function $\text{q}_l^{\pi_l}$ in (\ref{eq:distSARSA}) is a table in which the number of fields is equal to the number of states 
multiplied by the number of actions.
However, in our case the number of states is infinite.
This means, infinitely many values of $\text{q}_l^{\pi_l}(S_i, p_{l,i})$ would need to be stored. 
Since such a table cannot be constructed, linear function approximation is used to handle the infinite 
number of states.
With linear function approximation, $\text{q}_l^{\pi_l}(S_i, p_{l,i})$ is represented as the linear combination of a set of $M$ feature functions
$\text{f}_{l,m}(S_i, p_{l,i}),\; m=1,...,M$ which map the state-action pair $(S_{i},p_{l,i})$ onto a feature value.
The proposed feature functions are explained in Section \ref{subsec.FF}. 
For a given pair $(S_{i},p_{l,i})$, the feature values are collected in the vector $\textbf{f}_l \in \mathbb{R}^{M\times1}$ and the contribution of each
feature is included in the vector of weights $\textbf{w}_l \in \mathbb{R}^{M\times1}$. The action-value function is approximated as 
\begin{equation}
  \text{q}_l^{\pi_l}(S_{i},p_{l,i}) \approx \hat{\text{q}_l}^{\pi_l}(S_{i},p_{l,i},\textbf{w}_l)=\textbf{f}_l^\text{T}\textbf{w}_l.
  \label{eq:Qapprox}
\end{equation}

When SARSA with linear function approximation is applied, the updates are performed on the weights because they control the contribution of each feature
function on $\hat{\text{q}}_l^{\pi_l}(S_{i},p_{l,i})$.
In every time interval, the vector $\textbf{w}_l$ is adjusted in the direction that reduces the error between ${\text{q}}_l^{\pi_l}(S_{i},p_{l,i})$ and 
$\hat{\text{q}_l}^{\pi_l}(S_{i},p_{l,i},\textbf{w}_l)$ following the gradient descent approach presented in \cite{Sutton1998}.
Considering the update for $\text{q}_l^{\pi_l}(S_i,p_{l,i})$ given in (\ref{eq:distSARSA}), we propose to update $\textbf{w}_l$ as 
\begin{align}
\textbf{w}_{l,i+1} = \textbf{w}_{l,i}+\text{max}& \left\{ 0,\alpha_i\left[R_i+\gamma\,\textbf{f}_l^{T}(S_{i+1},p_{l,i+1})\textbf{w}_{l,i} 
			    -\textbf{f}_l^\text{T}(S_i,p_{l,i})\textbf{w}_{l,i}\right]\textbf{f}_l(S_i,p_{l,1})\right\}.
\label{eq:wUpdate}
\end{align}

\subsection{Partially observable states}
\label{subsec.partiallyObs}
So far, the fact that $S_i$ is only partially observable by the EH nodes was not taken into account. Here, we explain the proposed mechanisms used to overcome this partial observability.
As $\text{N}_1$ and $\text{N}_2$ have outdated channel state information, first, a channel predictor based on a Kalman filter is introduced at each of them to predict
their own channel coefficient $h_{1,i}$ and $h_{2,i}$, respectively.
Then, we propose to introduce a signaling phase in which $\text{N}_1$ and $\text{N}_2$ exchange the current values of their parameters.
\subsubsection{Channel predictor}
\label{subsec.kalman}
To predict the channel coefficient, a known symbol $s_{l,i}$ is assumed to be transmitted from $\text{N}_l$ to $\text{N}_{l+1}$. The received signal $y_{l+1,i}$ at $\text{N}_{l+1}$ in the low-pass domain is
\begin{equation}
y_{l+1,i}=s_{l,i}h_{l,i}+w_{l+1,i},
\label{eq:pilot}
\end{equation}
where $w_{l+1,i}$ accounts for the receiver noise and interference, and has variance $\sigma^2$.
Furthermore, the channel gain, $g_{l,i}=|h_{l,i}|$, is assumed to follow a Rayleigh distribution and the Jakes' model
\cite{Jakes74} is used to model the autocorrelation function ACF of the complex valued channel coefficients as
\begin{equation}
 \text{ACF}={J}_o(2 \pi f_{D,\text{max}}\tau),
 \label{eq:ACF}
\end{equation}  
where $J_o$ is the zero\textsuperscript{th} order Bessel function of the first kind and $f_{D,\mathrm{max}}$ is the maximum Doppler frequency. As 
extensively done in literature \cite{Shikur15, Chen04, McGuire05}, at each $\text{N}_l$, the dynamics of the channel coefficient are modeled as an 
autoregressive process of order $d$ and parameters $a_{l,1},...,a_{l,d}, \zeta $ as 
\begin{equation}
 h_{l,i}=-\sum_{j=1}^{d}{a_j h_{l,i-j}}+\zeta_l z_{l,i},
 \label{eq:ARprocess}
\end{equation}  
where $z_{l,i}$ is AWGN. The parameters $a_{l,1},...,a_{l,d}, \zeta_l$ are calculated at $\text{N}_l$ by means of solving the 
Yule-Walker equation considering the ACF in (\ref{eq:ACF}).
From (\ref{eq:pilot}) and (\ref{eq:ARprocess}), the state-space model for $h_{l,i}$ can be built. For this purpose, let us define the vectors 
$\mathbf{x}_{l,i}=[h_{l,i},h_{l,i-1},...,h_{l,i-d+1}]^\text{T}$, $\mathbf{g}_{l}=[\zeta_l,0,...,0]$ and $\mathbf{s}_{l,i}=[s_{l,i},0,...,0]$ such that 
\begin{align}
 \mathbf{x}_{l,i}=\mathbf{C}_l\mathbf{x}_{l,i-1}+\mathbf{g}_{l}\mathbf{v}_{l,i},\\
 y_{l+1,i}=\mathbf{s}_{l,i}\mathbf{x}_{l,i}+w_{l+1,i}
 \label{eq:stateSpace}
\end{align}  
where $\mathbf{v}_{l,i}$ is the white Gaussian process noise and
\begin{equation}
 \mathbf{C}_l = 
 \begin{pmatrix}
  -a_{l,1} & -a_{l,2} & \cdots & -a_{l,d} \\
         1 &        0 & \cdots &        0 \\
   \vdots  &   \vdots & \ddots & \vdots  \\
         0 &   \cdots &      1 & 0 
 \end{pmatrix}.
\end{equation} 
Using (\ref{eq:stateSpace}), each $\text{N}_l$ can estimate its own channel gain at time interval $i$ following the procedure described in Algorithm 
\ref{alg:kalmanFilter}.
The estimate $\hat{h}_{l,i}$  of the channel coefficient of $\text{N}_l$ in time interval $i$ is given by $\hat{h}_{l,i}=[1,0,...,0]\mathbf{x}_{l,i}$.
\begin{algorithm}[t]
\footnotesize
\begin{algorithmic}[1]
  \State initialize $\mathbf{x}_{l,1}=\mathbf{0}_d$ and $\mathbf{M}_{l,1}=\mathbf{I}_d$
    \For {every time interval $i=1,...,I$}
      \State set $\mathbf{M}_{l,i}=\mathbf{C}_{l}\mathbf{M}_{l,i-1} \mathbf{C}_l^\mathrm{H}+\mathbf{g}_{l} \mathbf{g}_{l}^\mathrm{H}$
      \State set $\psi = \mathbf{s}_{l,i}\mathbf{M}_{l,i}\mathbf{s}^\mathrm{H}_{l,i} + \sigma^2$
      \State calculate the Kalman gain $\mathbf{K}_{l,i}=\mathbf{M}_{l,i}\mathbf{s}^\mathrm{H}_{l,i}/\psi$
      \State update $\mathbf{x}_{l,i}=\mathbf{C}_{l}\mathbf{x}_{l,i-1} +\mathbf{K}_{l,i}(
      y_{l,i}-\mathbf{s}_{l,i}\mathbf{C}_{l,i}\mathbf{x}_{l,i-1})$
      \State update $\mathbf{M}_{l,i}=(\mathbf{I}_d-\mathbf{K}_{l,i}\mathbf{s}_{l,i})\mathbf{M}_{l,i}$      
      \State obtain $\hat{h}_{l,i}=[1,0,...,0]\mathbf{x}_{l,i}$
    \EndFor
\end{algorithmic}
\caption{Kalman filter based channel predictor } \label{alg:kalmanFilter}
\end{algorithm}

\subsubsection{Signaling}
\label{subsec.signaling}
We consider a transmission scheme which consists of a signaling phase and a data transmission phase. 
During the signaling phase of duration $\tau_\text{sig}$, the EH nodes exchange information about their current parameters, i.e., $\text{N}_1$ transmits $\{E_{1,i}, B_{1,i}$,$ D_{1,i}\}$ and $\text{N}_2$ transmits $\{E_{2,i}, B_{2,i}$, $ \hat{g}_{2,i}, D_{2,i}\}$.
$\text{N}_1$ does not transmit $\hat{g}_{1,i}$ because $g_{1,i}$ is already known at $\text{N}_2$.
During the data transmission phase of duration $\tau_\text{data}$, the EH nodes transmit the data stored in their data buffers. 
To facilitate the coordination among the nodes, we keep $\tau_\text{sig}$ fixed and in each time interval $i$, we calculate the power $p_{\text{sig},l,i}$ 
required for the transmission of the signaling. 
In the following, we describe how to compute $p_{\text{sig},l,i}$.

Let $x_{l,i}$ be a variable that represents any parameter of $\text{N}_l$, i.e., $x_{l,i} \in \{E_{l,i}, B_{l,i}, \hat{g}_{l,i}, 
D_{l,i}\}$. 
Then, the number  $L_{x_{l,i}}$ of bits required for the transmission of each $x_{l,i}$ depends on the type of quantizer that is used. For 
simplicity, in this paper we consider a uniform quantizer. Consequently, $x_{l,i}$ depends on the tolerable quantization error $e_{x_{l,i},\text{quant}}$,  
the maximum value $V_{x_{l,i},\text{max}}$ and the minimum value $V_{x_{l,i},\text{min}}$ each of them can take.
$L_{x_{l,i}}$ is calculated as
\begin{equation}
 L_{x_{l,i}}=\left\lceil\log_2\left(\frac{V_{x_{l,i},\text{max}}-V_{x_{l,i},\text{min}}}{e_{x_{l,i},\text{quant}}}\right)-1\right\rceil,
 \label{eq:noBits}
\end{equation}
where $\lceil \cdot \rceil$ is the rounding operation to the next integer value greater than or equal to the evaluated number. 
Since $V_{x_{l,i},\text{max}}$ and $V_{x_{l,i},\text{min}}$ are fixed for each $x_{l,i}$, the number of bits required for signaling is constant for all the time intervals and it is given by
\begin{equation}
L_l = \sum_{\forall x_{l,i}}{L_{x_{l,i}}}.
\end{equation}
Given $L_l$, the power $p_{\text{sig},l,i}$ required to transmit the signaling from $\text{N}_l$ to $\text{N}_j$ is 
\begin{equation}
p_{\text{sig},l,i}=\frac{\sigma^2}{|g_{l,i}|^2}\left( 2^{\frac{L_l}{W \tau_\text{sig}}}-1\right).
\label{eq:psig}
\end{equation}

It should be noted that the amount of energy $\tau_\text{sig} p_{\text{sig},l,i}$ used by each node for the transmission during the signaling phase is deducted from the battery level $B_{l,i}$ and the rest is available for data transmission. 
Moreover, if for any of the EH nodes the energy in the battery is lower than the value required to send the signaling and the tolerable quantization error is fixed, then the number of parameters sent during the signaling phase is reduced. 
The order in which this reduction is done is given by the impact each parameter has on the feature functions described in Section \ref{subsec.FF}.
First, the transmission of $E_{l,i}$ is skipped.
If the energy in the battery is not sufficient, then the transmission of $D_{l,i}$ is skipped as well. Finally, if the energy is still not sufficient, also the transmission of $B_{l,i}$ is skipped.
When $\text{N}_l$ cannot transmit the signaling, $\text{N}_j, \, j\in\{1,2\},\,j\neq l$, assumes that $\text{N}_l$ has harvested an amount of energy equal to its own, i.e., $E_{l,i}=E_{j,i}$, and that the signaling was not sent because the battery level of $\text{N}_l$ is zero, i.e., $B_{l,i}=0$.
Additionally, since there is no knowledge about the channel gain, it is assumed that $\hat{g}_{l,i}=\hat{g}_{l,i-1}$. 
For the data buffer level of node $\text{N}_l$, it is assumed that $D_{l,i}=\text{max}\{0,D_{l,i-1}-R_{l,i-1}\}$, where $R_{l,i}$ is the number of bits 
transmitted by $\text{N}_l$ in 
time interval $i-1$.
The overhead caused by the transmission during the signaling phase is measured in bits. 
As described in (\ref{eq:throughput}), this overhead is not included in the calculation of the achieved throughput.

\subsection{Feature functions}
\label{subsec.FF}
The feature functions are generally defined based on the natural attributes of the problem which are in this case the EH processes at the EH nodes, the finite 
batteries, the data arrival processes, the finite data buffers and the channel fading processes.
For the proposed MARL, we consider $M=6$ binary feature functions.
Moreover, we extend our previous work \cite{Ortiz2016, Ortiz20162hop} in order to include the energy consumed by the circuit.
Each $\text{N}_l$ calculates the value of the feature functions based on its own parameters and the information obtained during the signaling phase.

The first feature function $\text{f}_1(S_{i},p_{l,i})$ considers the energy causality and battery overflow constraints.
It indicates whether in state $S_{i}$, a given $p_{l,i}$ avoids a battery overflow situation. Additionally, it indicates 
whether $p_{l,i}$ fulfills the energy causality constraint and it is written as
\begin{equation}
\text{f}_1(S_{i}, p_{l,i})=
  \begin{cases} 
   1, & \text{if } (B_{l,i}+E_{l,i}-E_\text{Circ}-\tau_{\text{data}} p_{l,i}\leq B_{\text{max},l}) \wedge (E_\text{Circ}+ \tau_\text{data} p_{l,i}\leq B_{l,i})\\
   0, & \text{else},
  \end{cases}
  \label{eq:f1}
\end{equation}
where $\wedge$ is the logical conjunction operation.

The second feature function $\text{f}_2(S_{i}, p_{l,i})$ considers the power allocation problem by performing water-filling between the current 
$\hat{g}_{l,i}$ and the mean value $\bar{g}_{l,i}$ of past channel realizations.
The water level $\upsilon_{l,i}$ is calculated as
  \begin{equation}
   \upsilon_{l,i}=\frac{1}{2}\left(\frac{B_{l,i}}{\tau_\text{data}}+\frac{E_{l,i}}{\tau_\text{data}}+\sigma^2\left(\frac{1}{|\bar 
g_{l,i}|}+\frac{1}{|\hat{g}_{l,i}|}\right)\right),
   \label{eq:waterlevel}
  \end{equation}
and the power allocation given by the water-filling algorithm is given by
  \begin{equation}
    p_{l,i}^{\text{WF}}=\min\left\{\frac{B_{l,i}}{\tau_\text{data}}, \max\left\{0,\upsilon_{l,i}-\frac{\sigma^2}{|\hat{g}_{l,i}|}\right\}\right\}.
    \label{eq:plWF}
  \end{equation}
As $p_{l,i}$ can only be selected from the discrete set $\mathcal{A}_l$, the computed $p_{l,i}^{\text{WF}}$ is rounded
such that $p_{l,i}^{\text{WF}}\in\mathcal{A}_l$ holds. 
$\text{f}_2(S_{l,i}, p_{l,i})$ is written in \cite{Ortiz2016} as
  \begin{equation}
    \text{f}_2(S_{i}, p_{l,i})=
    \begin{cases}
     1, & \text{if } p_{l,i}^{\text{WF}}=p_{l,i} \\
     0, & \text{else}.
    \end{cases}
    \label{eq:f2}
  \end{equation}

The third feature function $\text{f}_3(S_{i}, p_{l,i})$ handles the case when the harvested energy is larger than the battery capacity, i.e., $E_{l,i}\geq 
B_{\text{max},l}$. 
In this case, the battery should be depleted in order to minimize the amount of energy that is lost due to overflow.
$\text{f}_3(S_{i}, p_{l,i})$ is given by
  \begin{equation}
    \text{f}_3(S_{i}, p_{l,i})=
    \begin{cases}
     1, & \text{if } \left( E_{l,i}\geq B_{\text{max},l} \right) \wedge \left( p_{l,i}=\delta \lfloor \frac{B_{l,i}-E_\text{Circ}}{\tau_\text{data}\delta} \rfloor \right) \\
     0, & \text{else}.
    \end{cases}
    \label{eq:f3}
  \end{equation}

The fourth feature function $\text{f}_4(S_i,p_{l,i})$ considers the data causality constraint.
Let us define $R_{l,i}^{(p_{l,i})}$ as the throughput that would be achieved if $p_{l,i}$ is selected. 
$\text{f}_4(S_{i}, p_{l,i})$ indicates if $R_{l,i}^{(p_{l,i})}$ fulfills the constraint given that there is enough energy in the 
battery to select it:
  \begin{equation}
    \text{f}_4(S_{i}, p_{l,i})=
    \begin{cases}
      1, & \text{if }  \left(R^{(p_{l,i})}_{l,i}\leq D_{l,i}\right) \wedge \left(B_{l,i}-E_\text{Circ}\geq \tau_\text{data}p_{l,i}\right) \\
      0, & \text{else}.
    \end{cases}
    \label{eq:f4}
  \end{equation}

The fifth feature function $\text{f}_5(S_i,p_{l,i})$ aims at the depletion of the data buffers as a preventive measure against data buffer overflows and it is defined as in \cite{Ortiz20162hop} as
  \begin{equation}
    \text{f}_5(S_{i}, p_{l,i})=
    \begin{cases}
      1, & \text{if }  p_{l,i}= \underset{p'_{l,i}\in \mathcal{A}_l}{\text{ argmin }} D_{l,i}-R^{(p'_{l,i})}_{l,i}\\  			    
      0, & \text{else}.
    \end{cases}
    \label{eq:f5}
  \end{equation}

The sixth feature function $\text{f}_6(S_{i}, p_{l,i})$ takes the available information $\text{N}_l$ has about $\text{N}_j$, 
$l,j\in\{1,2\}$, $l\neq 
j$ into consideration and it is used to further avoid data buffer overflows at $\text{N}_2$.
We focus on the data buffer overflow of $\text{N}_2$ because the data buffer level $D_{2,i}$ depends on the throughput of $\text{N}_1$ and 
$\text{N}_2$.
On the contrary, the data buffer level at $\text{N}_1$ depends only on the throughput of $\text{N}_1$ and its data arrival 
process which we cannot control.
To avoid data buffer overflows at $\text{N}_2$, each $\text{N}_l$ determines an estimate of the power $\bar{p}_{j,i}$ 
to be selected by $\text{N}_j$, $l\neq j$ using the water-filling procedure in (\ref{eq:waterlevel}), (\ref{eq:plWF}) 
and (\ref{eq:f2}). 
With $\bar{p}_{j,i}$, the corresponding throughput $R_{j,i}^{(\bar{p}_{j,i})}$ is calculated and it is compared to the data buffer level $D_{j,i}$.
If $R_{j,i}^{(\bar{p}_{j,i})}>D_{j,i}$, then $\bar{p}_{j,i}$ is scaled down to the minimum power value $\bar{p}_{j,i} \in \mathcal{A}_j$ that can be used to 
deplete the data buffer at $\text{N}_j$.
The feature function is then defined for $l=1$ as
  \begin{equation}
    \text{f}_6(S_{i}, p_{l,i})=
    \begin{cases}
      1, & \text{if }  \left(R^{(p_{l,i})}_{l,i} + D_{2,i} - R_{j,i}^{(\bar{p}_{j,i})}\leq D_{\text{max},2} \right )  \wedge \left(R^{(p_{l,i})}_{l,i} + 
D_{2,i} - R_{j,i}^{(\bar{p}_{j,i})}\geq 0\right) \\
      0, & \text{else}.
    \end{cases}
    \label{eq:f6}
  \end{equation}  
In the case $l=2$, the indices $l$ and $j$ should be interchanged.

\subsection{Summary of the proposed MARL}
\label{subsec.summary}
\begin{algorithm}[t]
\footnotesize
\begin{algorithmic}[1]
  \State initialize $\gamma, \alpha, \epsilon$ and $\textbf{w}_l$
  \State estimate own channel coefficient, exchange parameters and observe state $S_{i}$ 
  \State select $p_{l,i}$ using the $\epsilon$-greedy policy	\Comment{Eq. \ref{eq:egreedy}}
     \For {every time interval $i=1,...,I$}
      \State transmit using the selected $p_{l,i}$ 
      \State calculate corresponding reward $R_{2,i}$ \Comment{Eq. (\ref{eq:throughput})}
      \State estimate own channel coefficient, exchange parameters and observe state $S_{i+1}$
      \State select next $p_{l,i+1}$ using the $\epsilon$-greedy policy \Comment{Eq. (\ref{eq:egreedy})}
      \State update $\textbf{w}_l$ \Comment{Eq. (\ref{eq:wUpdate})}
      \State set $S_{i}=S_{i+1}$ and $p_{l,i}=p_{l,i+1}$      
    \EndFor
\end{algorithmic}
\caption{Proposed MARL algorithm } \label{alg:ApproxSARSA_MA}
\end{algorithm}
The proposed MARL algorithm is summarized in Algorithm \ref{alg:ApproxSARSA_MA}.
First, each $\text{N}_l$ initializes the values for the discount factor $\gamma$, the learning rate $\alpha$, and the probability 
$\epsilon$.
Then, the EH nodes estimate their own channel coefficients, exchange their local causal knowledge during $\tau_\text{sig}$ and observe 
$S_i$. 
According to $S_i$ and using the $\epsilon-$greedy policy, each node selects its own $p_{l,i}$. 
After the data transmission phase, the nodes calculate the reward, estimate their own new channel coefficients, exchange their causal knowledge during a new signaling phase and observe the new state $S_{i+1}$.
Each node selects the new $p_{l,i+1}$ using the $\epsilon-$greedy policy and updates  its weights $\textbf{w}_l$.
The same procedure is repeated in every time interval.
\subsection{Complexity analysis}
\label{subsec.complexity}
The proposed MARL is an iterative algorithm. 
Therefore, to determine its computational complexity, only one iteration has to be considered.
By examining Algorithm \ref{alg:ApproxSARSA_MA}, it is clear that the computationally most demanding tasks are the estimation of the channel coefficient (Lines 2 and 7), the selection of the transmit power $p_{l,i}$ (Lines 3 and 8) and the update of $\mathbf{w}_l$ (Line 9). 
The complexity of the Kalman-filter based channel estimator scales as $O(d^{3})$ \cite{Daum2005}, where $d$ is the order of the filter. 
Furthermore, for the selection of $p_{l,i}$, the $\epsilon$-greeedy policy is considered.
In this case, the highest complexity is due to the calculation of $\text{Q}(S_i, p_{l,i})$ for all the possible actions and the selection of the $p_{l,i}$ that leads to the maximum $\text{Q}$. The computational complexity for the calculation of $\text{Q}(S_i, p_{l,i})$ is $O(|\mathcal{A}|M)$ while the selection of the maximum value scales as $O(|\mathcal{A}|)$.
Lastly, the update of $\mathbf{w}_l$ using (\ref{eq:wUpdate}) has a complexity of $O(M^2)$.
As in our model $d$ is fixed, 
the computational complexity of one iteration of the algorithm scales linearly with $|\mathcal{A}|$ and polynomially with $M$ as $O(2|A|M + M^2)$.
Following a similar procedure, we can determine that the computational complexity of the hasty policy, which does not perform any learning but aims at depleting the batteries in each time interval, scales as $O(|\mathcal{A}|)$. 
In our proposed MARL $M=6$. 
Furthermore, usually $|A|>>M$, e.g., $|A|\approx 100$. 
This means, the main contribution to the computational complexity of our proposed MARL is due to the dimensions of the action space and the extra complexity caused by the use of the linear function approximation is the price to be paid for the improvement in the performance.
An additional advantage of the iterative nature of our proposed MARL is that it reduces the memory requirements on the system. 
Note that even though a continuous state is considered, the use of linear function approximation causes that only the vector of weights needs to be stored in addition to the vector of features used to describe the state at time interval $i$.

 \section{Convergence Guarantees}
 \label{sec.Convergence}
 
 In this section, we provide convergence guarantees for the proposed MARL algorithm for the case when the EH nodes are able to perfectly observe the current system's state, i.e., when the signaling is successfully sent.
Furthermore, as the EH, data arrival and channel fading processes might be non-stationary, we consider a constant learning rate to ensure that the new obtained rewards are considered in the learning process \cite{Sutton1998}.
 Inspired by the work of \cite{Lauer2000}, we first show that the local action-value function $\text{q}_{l}^{\pi_l}(S_i, p_{l,i})$ is a projection of  
 the centralized action value function $\text{Q}^\Pi(S_i, P_i)$ that leads to the selection of the best action in $\text{Q}^\Pi(S_i, P_i)$ for $\text{N}_l$.
 As this proof only holds for the case of a finite number of states, we then show that the guarantees given in \cite{Gordon01} for the  update of weights in the single-agent RL algorithm SARSA with  linear function approximation holds for the proposed updating rule given in (\ref{eq:wUpdate}) for the multi-agent case.
 \cite{Gordon01} shows that in SARSA with linear function approximation, the weights converge to a bounded region.
 
 \begin{remark}
 For an $n$-player Markov game defined by the  $(\mathcal{S}, \mathcal{A}_1, ..., \mathcal{A}_n, \mathcal{T}, \mathcal{R}_1, ..., \mathcal{R}_n)$ in 
which 
 the nodes have the same reward function $\mathcal{R}_1=...=\mathcal{R}_n=\mathcal{R}$, $\mathcal{R}\geq 0$, the equality 
   \begin{equation}
     \text{q}_{l,i}(S_i,p_{l,i}) = \underset{\substack{P_i=(p_{1,i},...,p_{n,i})\\P^{(l)}_i=p_{l,i}}}{\mathrm{max}} \text{Q}_i(S_i,P_i),
     \label{eq:proposition}
   \end{equation}
 where $\text{Q}_i(S_i,P_i)$ and $\text{q}_{l,i}(S_i,p_{l,i})$ are the values of the centralized and local action-value function in time interval 
 $i$, respectively, holds for any player $l$, any $S_i$, and any individual action $p_{l,i}$ in time interval $i$. 
 The values of $\text{Q}_i(S_i,P_i)$ and $\text{q}_{l,i}(S_i,p_{l,i})$ are updated in each time interval using (\ref{eq:centralizedSARSA}) and 
 (\ref{eq:distSARSA}), respectively, and by considering $\alpha=1$. 
 Additionally, $P^{(l)}_i$ is the $l^\text{th}$ element in $P_i$ corresponds to the action of player $l$ in time interval $i$ according to the 
centralized policy  $\Pi$. 
 \end{remark}
 
 \begin{proof}
 As in \cite{Lauer2000}, the proof is done by induction on $i$. 
 At $i=1$, no reward has been obtained.
 Therefore, $\text{Q}$ and $\text{q}_{l}$ are zero for every state $S_1 \in \mathcal{S}$ and $p_{l,1}\in \mathcal{A}_l$, $l\in\{1,...,n\}$ and 
(\ref{eq:proposition}) holds. 
 For arbitrary $i$, (\ref{eq:proposition}) holds for any pair  $(S_j,p_{k,j})$, $S_j\neq S_i$, $p_{k,j}\neq p_{l,i}$ because the updates in 
(\ref{eq:centralizedSARSA}) and (\ref{eq:distSARSA}) are only performed on the particular pair $(S_i, p_{l,i})$.
 Now, to prove (\ref{eq:proposition}) for the pair $(S_i, p_{l,i})$, we include the right side of (\ref{eq:proposition}) in the update of $\text{q}_{l,i}(S_i, 
p_{l,i})$ in (\ref{eq:distSARSA}) as
 \begin{align} 
   \text{q}_{l,i+1}(S_{i},p_{l,i}) = \mathrm{max} &   \left\{ \underset{\substack{P_i\\ P^{(l)}_i=p_{l,i}}}{\mathrm{max}}\text{Q}_i(S_i,P_i), \: 	
   R_i+\gamma \underset{\substack{P_{i+1}}}{\mathrm{max}}\:\text{Q}_i(S_{i+1},P_{i+1})\right\}.
  \label{eq:step1}
 \end{align}
By considering the equality $\text{max}(f(x)+a)=a+\text{max}f(x)$, (\ref{eq:step1}) can be rewritten as
\begin{align}
     \text{q}_{l,i+1}(S_{i},p_{l,i}) = \mathrm{max} &   \left\{ \underset{\substack{P_i\\P^{(l)}_i=p_{l,i}}}{\mathrm{max}} \text{Q}_i(S_i,P_i),  \:		
					\underset{\substack{P_{i+1}}}{\mathrm{max}} \Big\{R_i +\gamma  \text{Q}_i(S_{i+1},P_{i+1})\Big\}\right\}	
 					\label{eq:step3}
 \end{align}
From (\ref{eq:centralizedSARSA}), it is clear that the second term on the right side of (\ref{eq:step3}) corresponds to the update for the centralized action-value function $\text{Q}_{i+1}(S_i, P_i)$, which is independent of $P_{i+1}$.
We can then rewrite (\ref{eq:step3}) as
\begin{align}
  \text{q}_{l,i+1}(S_{i},p_{l,i}) = \mathrm{max} &   \left\{ \underset{\substack{P_i\\P^{(l)}_i=p_{l,i}}}{\mathrm{max}} \text{Q}_i(S_i,P_i),  \:		
					\text{Q}_{i+1}(S_{i},P_{i})\right\}	
 \label{eq:step3.5}
 \end{align}
 By expanding the term on the right side of (\ref{eq:step3.5}) we obtain
 \begin{align} 
  \text{q}_{l,i+1}(S_{i},p_{l,i}) = &   \mathrm{max}\left\{\left\{\text{Q}_i(S_i,P_i)\mid P^{(l)}_i=p_{l,i}, P_j\neq P_i\right\} \cup \right. \nonumber \\
 						&  \left. \qquad \: \; \left\{\text{Q}_{i}(S_i,P_i)\mid P^{(l)}_i=p_{l,i}, P_j=P_i\right\} \cup \big\{\text{Q}_{i+1}(S_i,P_i)\big\}\right\}.
 \label{eq:step4}						
 \end{align}
 The first term on the right side of (\ref{eq:step4}) is equal to $\text{Q}_{i+1}(S_i,P_i)$ because for $P_j\neq P_i$ there is no update.
 The second term is always smaller than or equal to 
 $\text{Q}_{i+1}(S_i, P_i)$ because $\text{Q}$ is monotonically increasing. 
 $\text{q}_{l,i}(S_{i},p_{l,i})$ is then written as
 \begin{align} 
  \text{q}_{l,i+1}(S_{i},p_{l,i}) = &  \mathrm{max}\left\{\left\{\text{Q}_{i+1}(S_i,P_i)\mid P^{(l)}_i=p_{l,i}, P_j\neq P_i\right\} \cup \left\{\text{Q}_{i+1}(S_i,P_i)\right\}\right\} \nonumber \\
  = & \underset{\substack{P_i=(p_{1,i},...,p_{n,i})\\P^{(l)}_i=p_{l,i}}}{\mathrm{max}} \text{Q}_{i+1}(S_i,P_i).
 \end{align}
 \end{proof}
 
 As mentioned before, the previous proof only applies to the case when a finite number of states is considered. 
 For an infinite number of states, we use linear function approximation and propose an update rule of the weights of the approximation in (\ref{eq:wUpdate}).
 In \cite{Gordon01}, the author proves that single-agent SARSA with linear function approximation converges to a region. 
 It is straightforward to see that the same proof applies to the proposed update in (\ref{eq:wUpdate}).
 The reasoning is as follows. From (\ref{eq:wUpdate}), it is clear that the weights are only updated if the condition
 \begin{align}
 \alpha_i\left[R_i+\gamma\,\textbf{f}_l^{T}(S_{i+1},p_{l,i+1})\textbf{w}_{l,i}-\textbf{f}_l^\text{T}(S_i,p_{l,i})\textbf{w}_{l,i}\right] 
 \textbf{f}_l(S_i,p_{l,1})>0
 \label{eq:condition}
 \end{align}
 is fulfilled. 
 In this case, the updating rule of the weights corresponds to the one for single-agent SARSA which is given by
 \begin{align}
  \textbf{w}_{l,i+1} = & \,\textbf{w}_{l,i} + \alpha_i \left[ R_{l,i}+\gamma \textbf{f}^{\text{T}}(S_{l,i+1},P_{l,i+1})\textbf{w}_{l,i}
   -{\textbf{f}}^\text{T}(S_{l,i},P_{l,i})\textbf{w}_{l,i})\right] \textbf{f}(S_{l,i},P_{l,i}).
  \label{eq:wUpdateSARSA}
 \end{align}
 If (\ref{eq:condition}) is not fulfilled, the weights are not updated in the time interval and $\textbf{w}_{l,i+1}=\textbf{w}_{l,i}$.
 
 
 \section{Simulation Results}
 \label{sec.Simulation}

In this section, we present numerical results for the evaluation of the proposed MARL algorithm.
For the simulations, we consider the variables listed in Table \ref{table:parameters} unless it is otherwise specified.
 \begin{table}
	\caption{Simulation set-up}
	 \centering{ 
	\begin{tabular}{lcc|lcc}
		\hline \hline
		Variable description & Name & Value & Variable description & Name & Value\\
		\hline
		Avg. SNR per link & $\eta$ & 5dB									&		Exploration probability & $\epsilon$ & $1/i$ \\	
		Avg. SNR pilot signals & $\eta_\text{pilot}$ & 5dB					&		Kalman filter order & $d$ & 2\\		
		Bandwidth & $W$ & 1 MHz 											&		Learning rate & $\alpha$ & $1/i$\\
		Battery size & $B_{\text{max},l}$ & $\varsigma E_{\text{max},l}$	&		Number of realizations & $T$ & 1000\\
		Battery size factor & $\varsigma$ & 5								&		Number of time intervals & $I$ & 1000\\ 
		Circuit power consumption & $p_\text{Cir}$ & 100mW					&		Power density EH source & $\rho$ & 10mW/cm$^2$\\
		Data buffer size $\text{N}_1$ & $D_{\text{max},1}$ & $\infty$		&		Quantization error & $e_{x_{l,i},\text{quant}}$ & 1\%\\
		Data buffer size $\text{N}_2$ & $D_{\text{max},2}$ & $W\tau \log_2(1+ \beta \eta)$ & Time interval duration  & $\tau$ & 10ms\\
		Data buffer size factor & $\beta$ & 1								&		Signaling phase duration & $\tau_\text{sig}$ & $0.01\tau$ \\
		Discount factor & $\gamma$ & 0.9									&		Size of EH panel & $A$ & $4\times 4$cm\\
		\hline \hline
	\end{tabular}}
	\label{table:parameters} 
\end{table}
 The energies $E_{l,i}$, $l\in\{1,2\}$, harvested by $\mathrm{N}_1$ and $\mathrm{N}_2$ in time interval $i$ are taken from a uniform distribution with maximum value $E_{\text{max},l}$ and the channel coefficients are modeled as complex Gaussian processes using the model described in \cite{Kuehne2011}.
 To compare the performance of our proposed MARL algorithm, we consider the following alternative approaches:
 \begin{itemize}
 	\item Offline optimum: Assuming that a central entity has perfect non-causal knowledge regarding the EH, data arrival and channel fading processes of $\text{N}_1$ and $\text{N}_2$, an optimization problem is formulated which maximizes the throughput.
 	To ensure the feasibility of the offline optimization problem, we assume $E_\text{circ}=0$ for this approach.
 	\item Dist. Learning - No Cooperation: It is assumed that the EH nodes have only 
causal knowledge regarding their own EH, data arrival and channel fading processes. 
 No cooperation between the EH nodes to exchange their parameters is assumed and the transmission policy is obtained by solving independent RL problems 
at $\text{N}_1$ and $\text{N}_2$.
	\item Centralized Learning: Using the signaling phase to observe the system state, a centralized RL problem is considered in which $\text{N}_2$ decides jointly on the transmit power of $\text{N}_1$ and $\text{N}_2$. Note that this approach also considers the use of Kalman filter based channel estimators at the nodes.
 	\item Hasty policy: Depletes the battery of $\text{N}_1$ in each time interval to transmit the maximum possible amount of data to $\text{N}_2$. 
 At $\text{N}_2$, the policy aims at depleting the data buffer by selecting the maximum transmit power value that fulfills the data causality constraint.
 \end{itemize}
\begin{figure}
    \centering
    \begin{subfigure}[b]{0.45\textwidth}
		\resizebox{\linewidth}{!}{\input{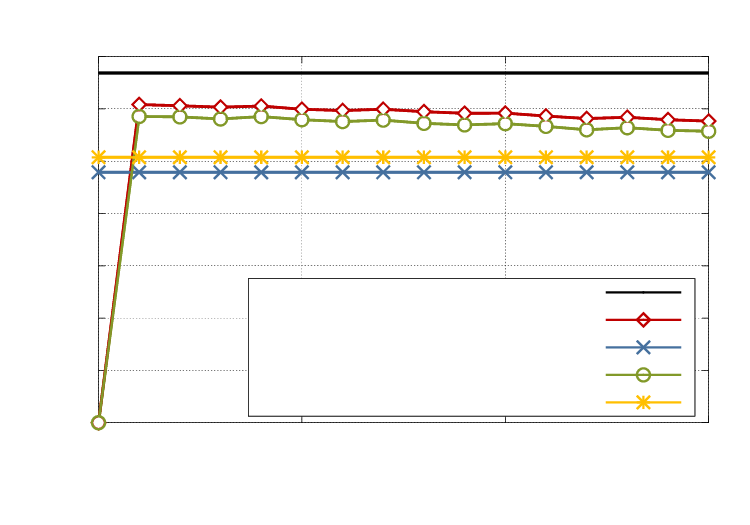}}
        \vspace{-10mm}
        \caption{$E_\text{Circ}=0$}
        \label{fig:tauSignoEcirc}
    \end{subfigure}
    ~ 
    \begin{subfigure}[b]{0.45\textwidth}
        \resizebox{\linewidth}{!}{\input{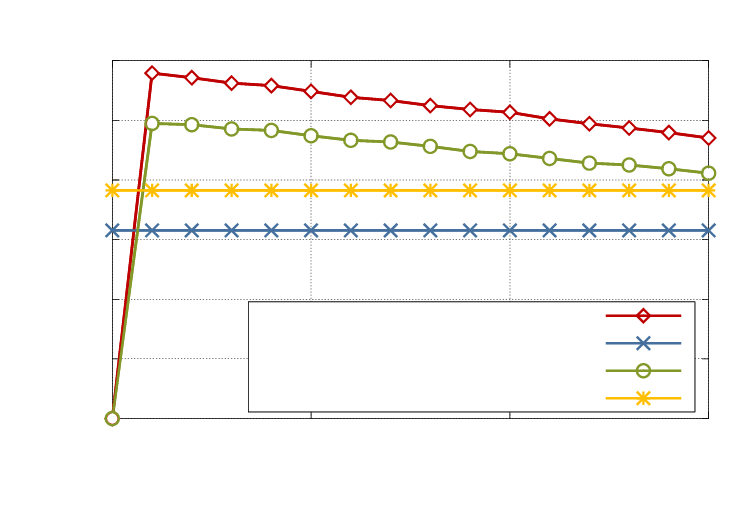}}
        \vspace{-10mm}
        \caption{$E_\text{Circ}=\tau p_\text{Circ}$}
        \label{fig:tauSigEcirc}
    \end{subfigure}
    \vspace{-4mm}
    \caption{Average throughput vs. fraction of time $\tau_\text{sig}/\tau$ assigned to signaling.}\label{fig:tauSig}
\end{figure}
%
%
In Fig. \ref{fig:tauSignoEcirc} and \ref{fig:tauSigEcirc}, we compare the average throughput, measured in bits, for different values of the fraction $\tau_\text{sig}/\tau$ of the duration of the time interval assigned to the signaling phase for $I=100$. 
Fig. \ref{fig:tauSignoEcirc} considers that $E_\text{Circ}=0$ and as expected, the largest throughput is achieved by the offline optimum policy which provides the upper bound of the performance assuming perfect non-causal knowledge of the system dynamics.
 Furthermore, the throughput achieved by the learning approach without cooperation and the hasty policy is flat because they do not consider a signaling phase and the complete duration of the time interval is used for the transmission of data.
 The achieved throughput of the proposed MARL and the centralized learning depends on the time assigned for the signaling.
 For $\tau_\text{sig}/\tau<15\%$, the proposed MARL outperforms the other approaches which also consider only causal knowledge.
 The reason for this improvement is that by including the signaling phase, $\text{N}_1$ and $\text{N}_2$ overcome the partial observability of the system state and are  able to learn a transmission policy that adapts to the battery levels, data buffer levels and channel gains of both nodes.
Moreover, the proposed MARL outperforms the centralized approach because in a distributed solution, a smaller action space needs to be considered, which increases the learning speed.
In the figure, the maximum throughput of the proposed MARL is achieved at $\tau_\text{sig}/\tau=1\%$.
For $\tau_\text{sig}/\tau<1\%$, the throughput is reduced because, as shown in (\ref{eq:psig}), the relation between $\tau_\text{sig}$ and $p_{\text{sig},l,i}$ required to transmit the signaling is not linear and the smaller $\tau_\text{sig}$, the over-proportionally larger 
 $p_{\text{sig},l,i}$. 
 As $p_{\text{sig},l,i}$ increases, the probability of not having enough energy in the battery to fulfill this requirement increases. Consequently, the nodes do not have enough energy to transmit during the signaling phase and to exchange their causal knowledge.
 When $\tau_\text{sig}/\tau$ increases to values beyond 1\%, the achieved throughput decreases.
 Even though for $\tau_\text{sig}/\tau>1\%$, the EH nodes have a longer signaling phase to exchange their causal knowledge, and can therefore use less 
power for the transmission of the signaling and save energy for data transmission, less time is left for the transmission of data. 
 As a result, the power required to transmit a certain amount of data increases.
In Fig. \ref{fig:tauSigEcirc} the energy $E_\text{Circ}$ consumed by the circuit is considered. 
In this case, the offline optimum is not included because for such scenario, the feasibility cannot be guaranteed.
When $E_\text{Circ}\neq 0$, the throughput of all the approaches is reduced because less energy is available for data transmission. 
However, the trend of their performance remains as in Fig.\ref{fig:tauSignoEcirc}.
For $\tau_\text{sig}/\tau=1\%$, the proposed MARL approach achieves a throughput which is {17}\% larger than for the centralized approach, {83}\% larger than for the distributed learning approach without cooperation and 51\% larger than for the hasty policy.

\begin{figure*}
\begin{minipage}{.475\textwidth}
	\centering
	\resizebox{\linewidth}{!}{\input{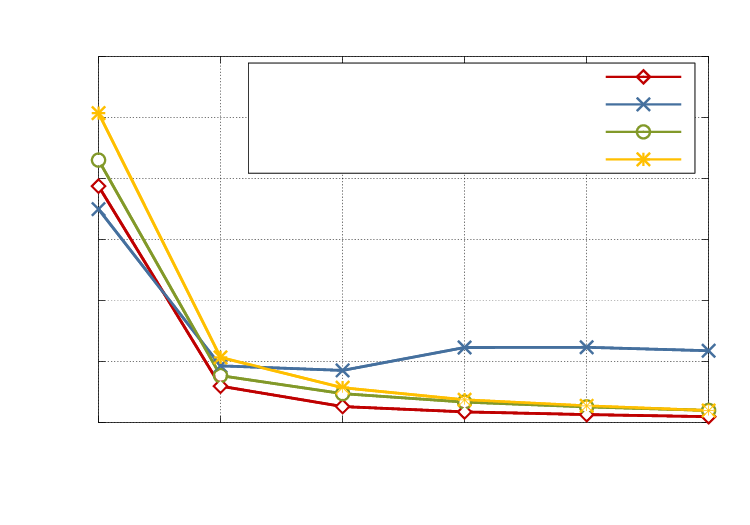}}
	\vspace{-12mm}
	\caption{Impact of data buffer size factor $\beta $.}
		\vspace{-4mm}
	\label{fig:bufferSize}
\end{minipage}
\hspace{5mm}
\begin{minipage}{.475\textwidth}
	\centering
	\resizebox{\linewidth}{!}{\input{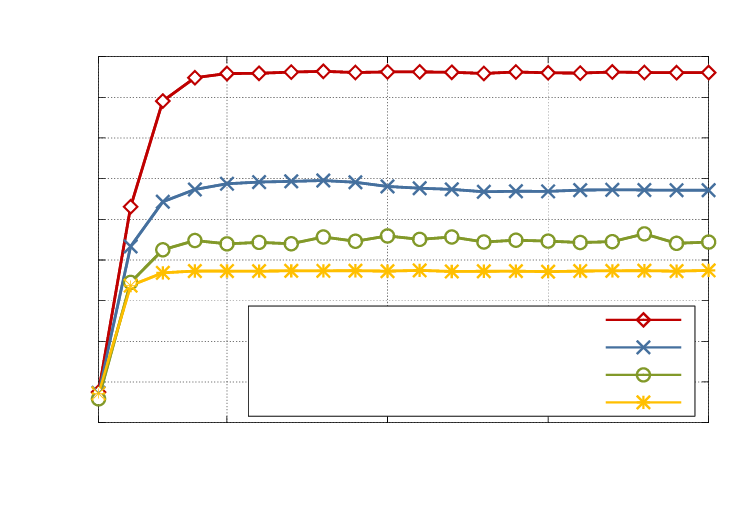}}
	\vspace{-12mm}
	\caption{Impact of data arrival process}
		\vspace{-4mm}
	\label{fig:packets}
\end{minipage}
\end{figure*}
The number of data buffer overflows at $\text{N}_2$ vs. the data buffer size is shown in Fig. \ref{fig:bufferSize}.
 We have omitted the result of the offline optimum because due to (\ref{eq:dataOverflow}), data buffer overflows are not allowed in the solution of 
the optimization problem{\footnote{ If data overflows cannot be avoided due to lack of energy, the optimization problem becomes infeasible.}.
 It can be seen that, as the data buffer size increases, the number of data buffer overflows is reduced for all the approaches, as expected.
 For $\beta=1$, the proposed MARL has 22\% less data buffer overflows than the centralized learning approach, 36\% less than the distributed learning approach without cooperation and 46\% less than the hasty policy. 
 The better performance of the proposed MARL results from the fact that by exchanging the causal knowledge during the signaling phase, $\text{N}_1$ knows the 
data buffer level of $\text{N}_2$ and can limit the amount of transmitted data when the data buffer of $\text{N}_2$ is almost full.
 It should be noted that although the proposed MARL is able to significantly reduce the number of data buffer overflows, it cannot reduce it to 
zero. 
 This is because non-causal knowledge is required to adapt the transmission policy according to the amounts of energy that will be harvested in 
 the future as well as the future channel gains.

Fig. \ref{fig:packets} shows the impact of the data arrival process at $\text{N}_1$. 
For this simulation, we consider that the number of data packets arriving in each time interval follows a Poisson distribution where the average number of 
data packets arriving per time interval is given by $\lambda$.
Moreover, we consider a packet size of 10 kbit.
 The offline optimum policy is not considered because the feasibility of the optimization problem depends on each particular realization of the data arrival 
process. 
 In Fig. \ref{fig:bufferSize}, it can be seen that for $\lambda=1$, all the approaches
achieve almost the same performance. 
 This is because for $\lambda=1$, the data buffer is almost empty all the time. 
 Therefore, data buffer overflows are unlikely and the data packets received by $\text{N}_1$ can be retransmitted by $\text{N}_2$ to $\text{N}_3$. 
 As the number of data packets received per time interval increases, our proposed MARL outperforms the centralized approach, the learning approach without cooperation and the hasty policy because it prevents data buffer overflows at $\text{N}_2$, as previously observed in Fig. \ref{fig:bufferSize}.
In this case, the performance of the centralized learning is further decreased because the consideration of the state of the data buffer at $N_\text{1}$ increases the dimensions of the state-action space.

The impact of the battery size on the achieved throughput is evaluated in Fig. \ref{fig:battery}.
Our proposed MARL outperforms the reference schemes when $B_{l,\text{max}}>E_{l,\text{max}}$.
For $\zeta=5$ it is able to achieve twice the throughput as compared to the distributed learning approach without cooperation. Moreover, its performance is 13\% and 47\% higher than for the centralized approach and for the hasty policy, respectively.
\begin{figure*}
\begin{minipage}{.475\textwidth}
	\centering{
	\resizebox{\linewidth}{!}{\input{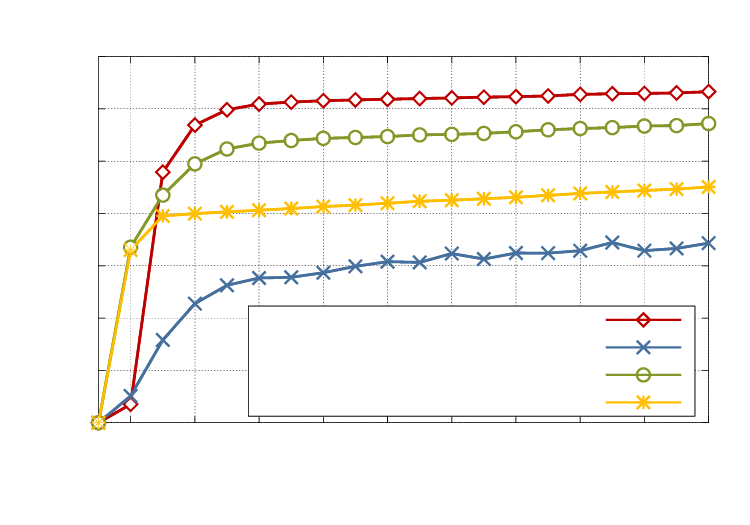}}
	\vspace{-12mm}
	\caption{Impact of the battery size.}
		\vspace{-4mm}
	\label{fig:battery}}
\end{minipage}
\hspace{5mm}
\begin{minipage}{.475\textwidth}
	\centering
    \resizebox{\linewidth}{!}{\input{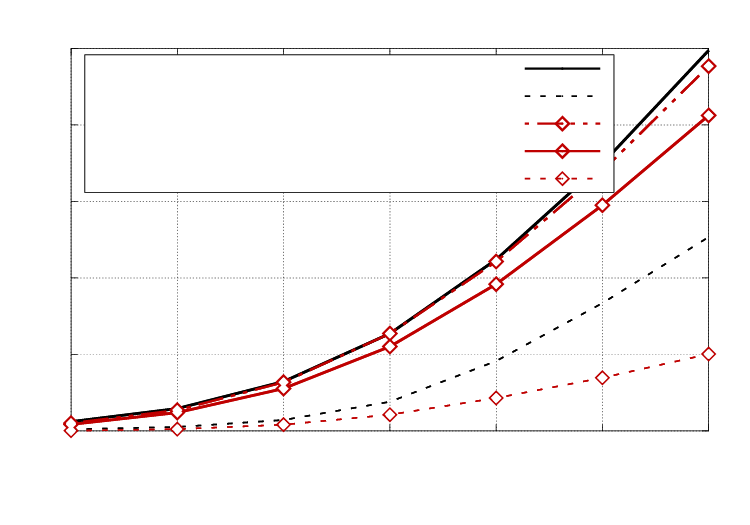}}
	\vspace{-12mm}
	\caption{Impact of average SNR per link}
		\vspace{-4mm}
	\label{fig:snr}
\end{minipage}
\end{figure*}
In Fig. \ref{fig:snr}, we compare the performance of the offline optimum policy and the proposed MARL for several values of the average SNR per link, i.e., from $\text{N}_1$ to $\text{N}_2$ and from $\text{N}_2$ to $\text{N}_3$.
To be able to calculate the throughput achieved by the offline optimum, $I=100$ time intervals and $E_\text{Circ}=0$ are considered.
We additionally evaluate the effect of the maximum amount of energy which $\text{N}_1$ and $\text{N}_2$ can harvest.
 For this purpose, we consider three different cases, i.e., $E_{\text{max},2}=10E_{\text{max},1}$, $E_{\text{max},2}=E_{\text{max},1}$ and 
$E_{\text{max},2}=0.1E_{\text{max},1}$.
 For the first case, i.e. $E_{\text{max},2}=10E_{\text{max},1}$, the offline optimum policy cannot be applied because battery overflows cannot be 
avoided at $\text{N}_2$ when it harvests much more energy than $\text{N}_1$.
 This is due to the fact that $\text{N}_2$ has more energy available in its battery than what is needed to retransmit the data it receives from $\text{N}_1$. 
 To allow battery overflows at $\text{N}_2$, a different optimization problem would need to be considered which is out of the scope of our work.
 In all the three cases, the throughput increases when the average SNR increases. 
 The largest throughput is achieved by the proposed MARL for the case when $E_{\text{max},2}=10E_{\text{max},1}$ and this throughput is close to the offline 
optimum performance for $E_{\text{max},2}=E_{\text{max},1}$.
 This is because harvesting more energy at $\text{N}_2$ cannot lead to a larger throughput if the amount of harvested energy is not  increased at $\text{N}_1$.
 The throughput is limited by the amount of data $\text{N}_1$ can transmit which in turn is limited by the amount of energy $\text{N}_1$ harvests, which for the 
two cases, $E_{\text{max},2}=10E_{\text{max},1}$ and $E_{\text{max},2}=E_{\text{max},1}$, is in a similar order of magnitude. 
 For $E_{\text{max},2}=E_{\text{max},1}$, the performance of the proposed MARL is reduced compared to the case when $E_{\text{max},2}=10E_{\text{max},1}$. 
 This is because there is less energy available at $\text{N}_2$. As a result, in each time interval, $\text{N}_2$ allocates less energy for data transmission.
 For the case when $E_{\text{max},2}=0.1E_{\text{max},1}$, the performance of the proposed MARL is close to the performance of the offline optimum policy in the low SNR regime, i.e., $\text{SNR}<10\text{dB}$. 
 This is due to the fact that in this case, $\text{N}_2$ is the bottleneck because it harvests on average much less energy than $\text{N}_1$. 
 Both approaches, the offline optimum policy and the proposed MARL, limit the amount of data $\text{N}_1$ transmits while aiming at maximizing the throughput in each time interval.

 Finally, in Fig. \ref{fig:Convergence}, we evaluate the convergence of the proposed MARL. 
 For this purpose, we compare the normalized throughput, i.e., the number of bits transmitted divided by the number $I$ of time intervals vs. the number $I$ 
of time intervals.
 In addition to the proposed MARL, the centralized approach and the distributed learning approach without cooperation, 
 we evaluate the performance of the proposed feature functions by implementing the proposed MARL using two standard approximation techniques, namely, fixed 
sparse representation (FSR) and radial basis functions (RBF) \cite{Geramifard2013}.
 Both, FSR and RBF are low-complexity techniques used to represent the continuous states. 
 For each $\text{N}_l$, $l\in\{1,2\}$, the state $S_i$, observed after the signaling phase, lies in an 8-dimensional space given by the parameters $E_{l,i}$, 
$B_{l,i}$, $g_{l,i}$ and $D_{l,i}$ of both nodes.
 In FSR, each dimension is split in tiles and a binary feature function is assigned to each tile. 
 A given feature function is equal to one if the corresponding variable is in the tile and zero otherwise \cite{Geramifard2013}.
 In our implementation, the tiles are generated by quantizing each dimension using the step size $\delta $ used in the definition of the action spaces $\mathcal{A}_l$.
 In RBF, each feature function has a Gaussian shape that depends on the distance between a given state and the center of the feature \cite{Sutton1998, Geramifard2013}.
 In contrast to FSR, in RBF a given state is represented by more than one feature function.
 In Fig. \ref{fig:Convergence}, it can be seen that the proposed MARL, the centralized approach and the distributed learning approach without cooperation converge at approximately the same number of iterations.
 This is due to the fact that the three approaches are based on the SARSA update. 
 However, since the proposed MARL considers the full cooperation among the EH nodes to exchange their causal knowledge, it can achieve a larger throughput.
 The number of feature functions required by a learning approach affects the performance because by increasing the number of feature functions used to represent the state space, a larger amount of weights has to be learned. 
 Consequently, our proposed MARL outperforms FSR and RBF because they require a larger number of feature functions compared to the proposed MARL which only needs six.
   \begin{figure}
  \vspace{-3mm}
  \centering
  	\resizebox{0.6\linewidth}{!}{\input{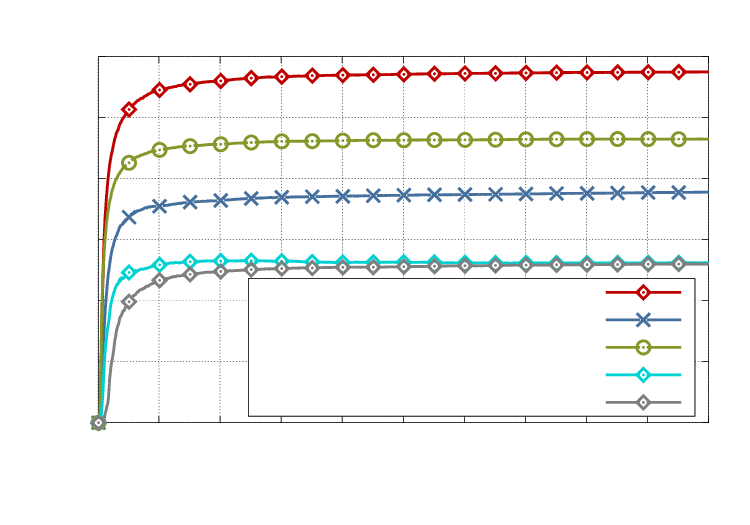}}
  \vspace{-6mm}
  \caption{Throughput vs. number $I$ of time intervals.}
  \label{fig:Convergence}
  \vspace{-6mm}
  \end{figure}
  
 To summarize the simulation results, it can be seen that with a proper selection of $\tau_\text{sig}$, the proposed MARL, which considers cooperation between the EH nodes, outperforms other approaches which also only consider causal knowledge but without cooperation between the nodes. 
 This means that reserving a fraction of time for the exchange of signaling among the nodes is more beneficial than assuming no cooperation at all, even though the time dedicated to data transmission is reduced in order to include the signaling phase. When the nodes cooperate with each other, a higher throughput can be achieved. 
 Furthermore, the proposed MARL reduces the number of data buffer overflows at $\text{N}_2$ as compared to the other approaches. This implies a reduction in the number of required retransmissions. 
 
 \section{Conclusion}
 \label{sec.Conclusion}
We have investigated an EH two-hop communication scenario where only partial causal knowledge regarding the EH processes, the data arrival processes and the channel fading processes was assumed at the EH transmitter and at the EH relay.
We considered the case when a signaling phase is available in each time interval. 
This signaling phase is used by the EH nodes to cooperate with each other by exchanging their own causal knowledge.
After the signaling phase, the EH nodes exploit the obtained knowledge to find  transmission policies which adapt to the battery levels, data buffer levels and 
channel gains of the EH nodes and which aim at maximizing the throughput. 
 We modeled the problem as a Markov game and proposed a multi-agent RL algorithm to find the transmission policies at the transmitter and at the relay.
 Furthermore, we have provided convergence guarantees for the proposed algorithm. 
 Through several simulation results we have shown that a larger throughput can be achieved when cooperation among the EH nodes is considered, compared to the case when no cooperation is assumed even after the signaling overhead is subtracted from the number of bits transmitted.
 Moreover, we have shown the trade-off between the duration of the signaling phase and the performance of the proposed algorithm and we have shown that the number of data buffer overflows is reduced when our proposed algorithm is considered.
 The distributed nature of our proposed algorithm makes it suitable for more complex relay networks, e.g., multi-hop networks.
%
%







%

\bibliographystyle{IEEEtran}
\bibliography{EH}

%


%
%




\end{document}